\documentclass[a4paper,12pt]{article}
\usepackage{amssymb}
\usepackage{amsmath}
\usepackage{amsthm}
\usepackage{latexsym}
\newcommand{\be}{\begin{eqnarray}}
\newcommand{\ee}{\end{eqnarray}}
\newcommand{\bdm}{\begin{displaymath}}
\newcommand{\edm}{\end{displaymath}}
\newtheorem{thm}{Proposition}
\newtheorem*{lm}{Lemma}

\begin{document}
\title{\Large\textbf{Automorphism Inducing Diffeomorphisms, Invariant
Characterization of Homogeneous 3-Spaces and Hamiltonian Dynamics
of Bianchi Cosmologies}}
\author{\textbf{T. Christodoulakis}\thanks{e-mail:
tchris@cc.uoa.gr}~~~\textbf{E. Korfiatis}~~\textbf{\& ~G. O.
Papadopoulos}\thanks{e-mail: gpapado@cc.uoa.gr}}
\date{}
\maketitle
\begin{center}
\textit{University of Athens, Physics Department\\
Nuclear \& Particle Physics Section\\
Panepistimioupolis, Ilisia GR 157--71, Athens, Hellas}
\end{center}
\vspace{1cm} \numberwithin{equation}{section}
\begin{abstract}
An invariant description of Bianchi Homogeneous (B.H.) 3-spaces is
presented, by considering the action of the Automorphism Group on
the configuration space of the real, symmetric, positive definite,
$3\times 3$ matrices. Thus, the gauge degrees of freedom are
removed and the remaining (gauge invariant) degrees, are the --up
to 3-- curvature invariants. An apparent discrepancy between this
Kinematics and the Quantum Hamiltonian Dynamics of the lower Class
A Bianchi Types, occurs due to the existence of the Outer
Automorphism Subgroup. This discrepancy is satisfactorily removed
by exploiting the quantum version of some classical integrals of
motion (conditional symmetries) which are recognized as
corresponding to the Outer Automorphisms.
\end{abstract}
\newpage
\section{Introduction}
In a preceding work \cite{chrissht} we have shown how the presence
of the linear constraints, entails a reduction of the degrees of
freedom for the quantum theory of Class A spatially homogeneous
geometries: the initial six-dimensional configuration space
spanned by $\gamma_{\alpha\beta}$'s (the components of the spatial
metric with respect to the invariant basis one-forms), is reduced
to a space parameterized by the independent solutions to the
linear quantum constraints (Kucha\v{r}'s physical variables
\cite{hajiceck}). For Bianchi Types $VI_{0}, VII_{0}, VIII, IX$
these solutions are the three combinations: \bdm
x^{1}=C^{\alpha}_{\mu\kappa}C^{\beta}_{\nu\lambda}\gamma_{\alpha\beta}
\gamma^{\mu\nu}\gamma^{\kappa\lambda}~~~~~
x^{2}=C^{\alpha}_{\beta\kappa}C^{\beta}_{\alpha\lambda}\gamma^{\kappa\lambda}~~~~~
x^{3}=\gamma \edm (or any other three, independent, functions
thereof) and the Wheeler-DeWitt equation becomes a P.D.E. in terms
of these $x^{i}$'s. The Bianchi Type $I$, where all structure
constants are zero (and thus the linear constraints vanish
identically), has been exhaustively treated \cite{ashtekar}. The
Type $II$ case, where only two linear constraints are independent,
has been examined along the above lines in \cite{christype2} and
differently in \cite{lidsey}.

The fact that the quantum theory (within each one of the above
mentioned Bianchi Types) forces us to consider as equivalent any
two points $\gamma^{(1)}_{\alpha\beta},
\gamma^{(2)}_{\alpha\beta}$ in the configuration space if they
form the same triplet $(x^{i})$, seems quite intriguing. It is the
purpose of the present work, to investigate in detail the reasons
for this grouping of the $\gamma_{\alpha\beta}$'s.\\
The paper is organized as follows:

Section 2 begins with a careful examination of the action of the
general coordinate transformations group on
$\gamma_{\alpha\beta}$. The demand that the diffeomorphisms must
preserve the manifest homogeneity of the 3-spaces, singles out a
particular set of those transformations which has a well defined,
non trivial action on $\gamma_{\alpha\beta}$; this action is then
proven to be nothing but the action of the automorphism group
corresponding to each
arbitrary but given Bianchi Type.\\
The differential description of these automorphic motions, is
achieved by identifying the vector fields on the configuration
space which, through their integral curves, induce these motions.
The importance of Automorphisms in the theory of Bianchi Type
Cosmologies, has been stressed in \cite{jantzen}.\\
Concluding this section, we prove the following: if (within a
particular albeit arbitrary Bianchi Type) two points
$\gamma^{(1)}_{\alpha\beta}, \gamma^{(2)}_{\alpha\beta}$ laying in
the configuration space, correspond to the same scalar
combinations of $C^{\alpha}_{\mu\nu}$ and $\gamma_{\alpha\beta}$,
then $\gamma^{(2)}_{\alpha\beta}=
\Lambda^{\mu}_{\alpha}\Lambda^{\nu}_{\beta}\gamma^{(1)}_{\mu\nu}$
where $\Lambda$ is an element of the corresponding Automorphism
group.

In section 3 we briefly recapitulate the essential features of the
quantum theory developed in \cite{chrissht}, and we compare the
purely kinematical results of the previous section, with the
ensuing Quantum Hamiltonian dynamics.\\
For the lower Bianchi Types, this comparison reveals an apparent
mismatch between the dynamics and the kinematics. The gap is
bridged through the notion of conditional symmetries
\cite{kucharconditional}, i.e. some linear in momenta, integrals
of motion; their quantum counterparts constrain $\Psi$ to be a
function of the geometry only.

Finally, some concluding remarks are included in the discussion.
\section{Automorphism Inducing Diffeomorphisms}
In this section we shall first relate the action of the
Automorphism group on $\gamma_{\alpha\beta}$, to the action
induced on it by the class of General Coordinate Transformations
(G.C.T.'s) which are subject to the restriction of preservation
manifest spatial homogeneity. To this end, consider the spatial
line element: \be \label{lineelement41}
ds^{2}=\gamma_{\alpha\beta}(t)\sigma^{\alpha}_{i}(x)
\sigma^{\beta}_{j}(x)dx^{i}dx^{j} \ee where
$\sigma^{\alpha}_{i}(x)dx^{i}$ are the invariant basis 1-forms, of
some given Bianchi Type.

The spatial homogeneity of this line element, is of course,
preserved under any G.C.T. of the form:

\be \label{transformationx} x^{i}\longrightarrow
\widetilde{x}^{i}=f^{i}(x) \ee

Under such a transformation, $ds^{2}$ simply becomes: \be
\label{lineelement42}
(ds^{2}\equiv)d\widetilde{s}^{2}=\gamma_{\alpha\beta}(t)
\widetilde{\sigma}^{\alpha}_{m}(\widetilde{x})\widetilde{\sigma}^{\beta}_{n}
(\widetilde{x})d\widetilde{x}^{m}d\widetilde{x}^{n} \ee where the
basis one-forms are supposed to transform in the usual way: \be
\label{sigmatransformation}
\widetilde{\sigma}^{\alpha}_{m}(\widetilde{x})=\sigma^{\alpha}_{i}(x)\frac{\partial
x^{i} }{\partial \widetilde{x}^{m} } \ee

If one were to stop at this point, then one might have concluded
that all spatial diffeomorphisms, act trivially on
$\gamma_{\alpha\beta}$ i.e. $\gamma_{\alpha\beta}\longrightarrow
\widetilde{\gamma}_{\alpha\beta}=\gamma_{\alpha\beta}$. But as we
shall immediately see, there are special G.C.T.'s which induce a
well-defined, non-trivial action on $\gamma_{\alpha\beta}$. To
uncover them, let us ask what is the change in form induced, by
tranformation (\ref{transformationx}), to the line element
(\ref{lineelement41}). To find this change we have to express the
line element (\ref{lineelement42}) in terms of the old basis
one-forms (at the new point) $\sigma^{\alpha}_{i}(\widetilde{x})$.
There is always a non singular matrix
$\Lambda^{\alpha}_{\beta}(\widetilde{x})$ connecting
$\widetilde{\sigma}$ and $\sigma$ i.e.: \be
\label{sigmatransformationp}
\widetilde{\sigma}^{\alpha}_{m}(\widetilde{x})=\Lambda^{\alpha}_{\mu}(\widetilde{x})
\sigma^{\mu}_{m}(\widetilde{x}) \ee Using this matrix $\Lambda$ we
can write line element (\ref{lineelement42}) in the form: \be
d\widetilde{s}^{2}=\gamma_{\alpha\beta}(t)\Lambda^{\alpha}_{\mu}(\widetilde{x})\Lambda^{\beta}_{\nu}
(\widetilde{x})\sigma^{\mu}_{m}(\widetilde{x})\sigma^{\nu}_{n}
(\widetilde{x})d\widetilde{x}^{m}d\widetilde{x}^{n} \ee

If the functions $f^{i}$, defining the transformation, are such
that the matrix $\Lambda^{\alpha}_{\mu}$ does not depend on the
spatial point, then there is a well defined, non trivial action of
these transformations on $\gamma_{\alpha\beta}$: \be \label{eq27}
\gamma_{\alpha\beta}\longrightarrow
\widetilde{\gamma}_{\mu\nu}=\Lambda^{\alpha}_{\mu}\Lambda^{\beta}_{\nu}\gamma_{\alpha\beta}
\ee With the use of (\ref{sigmatransformation}) and
(\ref{sigmatransformationp}), the requirement that
$\Lambda^{\alpha}_{\mu}$ does not depend on the spatial point
$\widetilde{x}^{i}$, places the following differential
restrictions on the $f^{i}$s: \be \label{fsigma} \frac{\partial
f^{i}(x)}{\partial
x^{j}}=\sigma^{i}_{\alpha}(f)S^{\alpha}_{\beta}\sigma^{\beta}_{j}(x)
\ee where $\sigma^{i}_{\alpha}$ and $S^{\alpha}_{\beta}$ are the
matrices inverse to $\sigma^{\alpha}_{i}(x)$ and
$\Lambda^{\alpha}_{\beta}$, respectively. These conditions
constitute a set of first order, highly non-linear P.D.E.'s in the
unknown functions $f^{i}$. The existence of solutions to these
equations, is guaranteed by the Frobenius theorem \cite{warner},
as long as the necessary and sufficient conditions
$\partial^{2}_{k,~l}f^{i}-\partial^{2}_{l,~k}f^{i}=0$ hold.
Through the use of (\ref{fsigma}) and the defining property of the
invariant basis 1-forms (\ref{2form}), we can transform these
conditions into the form: \be
2\sigma^{i}_{\alpha}(f)\sigma^{\epsilon}_{k}(x)\sigma^{\delta}_{l}(x)
\left(C^{\rho}_{\epsilon\delta}S^{\alpha}_{\rho}-C^{\alpha}_{\mu\nu}
S^{\mu}_{\epsilon}S^{\nu}_{\delta}\right) =0 \ee which is
satisfied, if and only if, $S^{\alpha}_{\mu}$ (and thus also
$\Lambda^{\alpha}_{\mu}$) is a Lie Algebra  Automorphism (see
\ref{automorphism} bellow). It is, therefore, appropriate to call
the General Coordinate Transformations (\ref{transformationx}),
when the $f^{i}$'s satisfy (\ref{fsigma}), \emph{Automorphism
Inducing Diffeomorphisms} (A.I.D.'s). The existence of such
spatial coordinate transformations is not entirely unexpected: in
the particular case
$\Lambda^{\alpha}_{\beta}(\widetilde{x})=\delta^{\alpha}_{\beta}$
these coordinate transformations, are nothing but the finite
motions induced on the hypersurface, by the three Killing vector
fields (existing by virtue of homogeneity of the space), which
leave the basis one-forms form invariant. The new thing we learn,
is that there are further motions leaving the basis one-forms
quasi-invariant i.e. invariant modulo a global (space independent)
linear mixing, with the mixing matrix $\Lambda^{\alpha}_{\mu}$,
belonging to the Automorphism Group. The notion of such
transformations `'leaving the invariant triads unchanged modulo a
global rotation`' also appears in Ashtekar's work \cite{ashtekar},
under the terminology `'Homogeneity Preserving Diffeomorphisms`';
also the term global is there used in the topological sense.

In order to gain a deeper understanding of the implications of the
above analysis as well as the consequences of the kinematics on
the dynamics, we have to carefully consider the configuration
space and the differential description of the changes (\ref{eq27})
induced on it, by the A.I.D.'s.

Let us begin with some propositions about the space of $3\times 3$
real, symmetric, (positive definite) matrices:
\begin{thm}
The set $\Sigma$ of all $3\times 3$ real, symmetric, matrices,
forms a vector subspace of $GL(3, \Re)$, and is thus endowed with
the structure of a six-dimensional  manifold.
\end{thm}
\begin{thm}
The set $\Delta$ of all $3\times 3$ real, symmetric, positive
definite, matrices, is an open subset of $\Sigma$.
\end{thm}
\begin{proof}
Let $\gamma_{\alpha\beta}$, be a positive definite $3\times 3$
real, symmetric, matrix, and \bdm p(s)=s^{3}-As^{2}+Bs-C \edm its
characteristic polynomial with $A, B, C$, continuous, polynomial,
functions of  $\gamma_{\alpha\beta}$'s. Since
$\gamma_{\alpha\beta}$ is symmetric, the necessary and sufficient
condition that $\gamma_{\alpha\beta}$ be positive definite, is
$A>0, B>0, C>0$. Therefore $\Delta$, as an inverse image of an
open subset, is itself open.
\end{proof}
\begin{thm}
The set $\Delta$ is an arcwise connected subset of $\Sigma$.
\end{thm}
\begin{proof}
Let $\gamma_{\alpha\beta} \in \Delta$. Then, there is $P \in
SO(3)$ such that (in matrix notation): \bdm P\gamma
P^{T}=D=\textrm{diag}(a,b,c), \edm with $a$, $b$, $c$ the three
positive eigenvalues of $\gamma$. Since $P$ belongs to $SO(3)$,
there is a continuous mapping $\omega:[0,1]\rightarrow SO(3)$ such
that $\omega(0)=P$ and $\omega(1)=I_{3}$. Introduce now the
mapping $f:[0,1]\rightarrow\Delta$, with
$f(\sigma)=\omega(\sigma)\gamma\omega(\sigma)^{T}$. As
$\omega(\sigma)$ belongs to $SO(3)$, its determinant is not zero
for every $\sigma \in [0,1]$. Therefore, by Sylvester's theorem,
$f(\sigma)$ is positive definite --just like $\gamma$. But
$f(0)=D$ and $f(1)=\gamma$, i.e. the matrix $\gamma$ is connected
to $D$, by a continuous curve lying entirely in $\Delta$. Consider
now the mapping: \bdm \phi:[0,1]\rightarrow\Delta \edm with: \bdm
\phi(\sigma)=\textrm{diag}((a-1)\sigma+1, (b-1)\sigma+1,
(c-1)\sigma+1) \edm $\phi$ is continuous and $\phi(\sigma) \in
\Delta,~\forall~\sigma \in [0,1]$. This means that $\gamma$ is
finally arcwise connected to $I_{3}$.
\end{proof}

\vspace{1cm} Let us now proceed with the differential description
of motions (\ref{eq27}). To this end, consider the following
linear vector fields defined on $\Delta$: \be \label{vectorfield}
X_{(i)}=\lambda^{\alpha}_{(i)\rho}\gamma_{\alpha\beta}\partial^{\beta\rho}\ee
with an obvious notation for the derivative with respect to
$\gamma_{\alpha\beta}$.\\
The matrices
$\lambda^{\beta}_{(i)\alpha}\equiv(C^{\beta}_{(\rho)\alpha},
\varepsilon^{\beta}_{(i)\alpha})$ are the generators of (the
connected to the identity component of) the Automorphism group
(see (\ref{generators})) and $(i)$ labels the different
generators. Depending on the particular Bianchi Type, the vector
fields (in $\Delta$) $X_{(i)}$ may also include, except of the
quantum linear constraints (generators of Inner Automorphic
Motions) $H_{\rho}=C^{\alpha}_{\rho\beta}\gamma_{\alpha\kappa}
\frac{\partial}{\partial\gamma_{\beta\kappa}}$, the generators of
the outer-automorphic motions:
$E_{(j)}\equiv\varepsilon^{\sigma}_{(j)\rho}\gamma_{\sigma\tau}\frac{\partial}{\partial\gamma_{\rho\tau}}$

The infinitesimal action of the generic vector field
(\ref{vectorfield}) $\varepsilon^{(i)}X_{(i)}$ on
$\gamma_{\alpha\beta}$ is given by: \be
\bar{\delta}\gamma_{\alpha\beta}\equiv
\varepsilon^{(i)}\frac{1}{2}(\lambda^{\mu}_{(i)\alpha}\gamma_{\mu\beta}+
\lambda^{\mu}_{(i)\beta}\gamma_{\mu\alpha}) \ee where
$\varepsilon^{(i)}$ are infinitesimal arbitrary parameters. If we
now define the matrices: \be \label{M}
M^{\mu}_{\alpha}=\varepsilon^{(i)}\lambda^{\mu}_{(i)\alpha} \ee we
can prove that these, are generators of automorphisms. To see it,
let us briefly recall the notion of a Lie Algebra Automorphism: if
$\mathcal{A}$ denotes the space of third rank (1,2) tensors under
$GL(3, \Re)$, antisymmetric in the two covariant indices, then the
structure constants transform (as it can be inferred from
(\ref{2form})) according to: \be
C^{\alpha}_{\mu\nu}\rightarrow\widetilde{C}^{\alpha}_{\mu\nu}=
S^{\alpha}_{\beta}\Lambda^{\kappa}_{\mu}\Lambda^{\lambda}_{\nu}
C^{\beta}_{\kappa\lambda} \ee with $\Lambda^{\alpha}_{\mu}$ and
$S^{\alpha}_{\mu}=(\Lambda^{-1})^{\alpha}_{\mu}$ $\in GL(3, \Re)$.
A transformation is called a Lie Algebra Automorphism, if and only
if, it leaves the structure constants unchanged i.e. if: \be
C^{\alpha}_{\mu\nu}=S^{\alpha}_{\beta}\Lambda^{\kappa}_{\mu}
\Lambda^{\lambda}_{\nu} C^{\beta}_{\kappa\lambda} \ee or
equivalently: \be \label{automorphism}
C^{\rho}_{\mu\nu}\Lambda^{\alpha}_{\rho}=\Lambda^{\kappa}_{\mu}
\Lambda^{\lambda}_{\nu} C^{\alpha}_{\kappa\lambda} \ee

To find the defining relation for the generators
$\lambda^{\alpha}_{\mu}$ of the automorphisms
$\Lambda^{\alpha}_{\mu}$, consider a path through the identity
$\Lambda^{\rho}_{\theta}(\tau)$, with
$\Lambda^{\rho}_{\theta}(0)=\delta^{\rho}_{\theta}$ (we are
concerned only with the connected to the identity component of the
automorphism group). Differentiating both sides of
(\ref{automorphism}) with respect to the parameter $\tau$ and
setting $\tau=0$, we get the relation: \be \label{generators}
\lambda^{\alpha}_{\beta}C^{\beta}_{\mu\nu}=\lambda^{\rho}_{\mu}
C^{\alpha}_{\rho\nu}+\lambda^{\rho}_{\nu}C^{\alpha}_{\mu\rho} \ee
where we have identified $\lambda^{\alpha}_{\mu}$'s with the
vectors tangent to the path, at the identity.

By virtue of the Jacobi Identities, one can see that a solution to
the system (\ref{generators}) is:
$\lambda^{\alpha}_{(\kappa)\beta}=C^{\alpha}_{(\kappa)\beta}$ and
thus, the structure constants matrices are the generators of the
\emph{Inner Automorphisms} proper invariant subgroup of $Aut(G)$.
For Bianchi Types $VIII, IX$ these, are the only generators of
automorphisms. For all other Bianchi Types, there exist extra
matrices satisfying (\ref{generators}) --say
$\varepsilon^{\alpha}_{(i)\beta}$-- generating the \emph{Outer
Automorphisms} subgroup of $Aut(G)$. We are now ready to find the
finite motions induced on $\Delta$, by the generic vector field
$X\equiv\varepsilon^{(i)}X_{(i)}$:
\begin{thm}
Let $\gamma^{(0)}_{\alpha\beta}$ be a fixed point in $\Delta$.
Then the curve $\gamma:\Re\rightarrow \Delta$ with: \bdm
\gamma_{\alpha\beta}(\tau)=(exp(\tau M))^{\mu}_{\alpha}(exp(\tau
M))^{\nu}_{\beta}\gamma^{(0)}_{\mu\nu} \edm is an integral curve
(passing through $\gamma^{(0)}_{\alpha\beta}$) of the vector field
$X\equiv\varepsilon^{(i)}X_{(i)}$.
\end{thm}
\begin{proof}
We give a rigorous proof of the statement that the matrices
$(exp(\tau M))^{\mu}_{\alpha}$, are automorphisms. To this end,
define the mapping
$\phi_{\tau}:\mathcal{A}\rightarrow\mathcal{A}$, with
$\phi_{\tau}(C)=\widetilde{C}$ where
$\widetilde{C}^{\alpha}_{\mu\nu}=S^{\alpha}_{\beta}
\Lambda^{\kappa}_{\mu}\Lambda^{\lambda}_{\nu}
C^{\beta}_{\kappa\lambda}$. Define also the matrices
$\Lambda(\tau)=exp(\tau M)$, $S(\tau)=exp(-\tau M)$, with $M$,
given by (\ref{M}). It is straightforward to verify that
$\phi_{\tau}\circ\phi_{\sigma}=\phi_{\tau+\sigma}$. Using the
Jacobi Identities and the definitions above, it is not difficult
to see that: \bdm \frac{d\phi_{\tau}(C)}{d\tau}|_{\tau=0}=0 \edm
Consider now two sets $\widetilde{C}, C \in \mathcal{A}$, such
that $\phi_{\psi}(C)=\widetilde{C}$, for some $\psi$. Since the
derivative of $\phi_{\theta}$ at 0 is zero, we have that: \bdm
\frac{d\phi_{\theta}(\widetilde{C})}{d\theta}|_{\theta=0}=\lim_{\theta
\to 0}\frac{\phi_{\theta}(\widetilde{C})-
\phi_{0}(\widetilde{C})}{\theta}=0 \edm which in turn, implies
that: \bdm \lim_{\theta \to
0}\frac{\phi_{\theta}\circ\phi_{\psi}(C)-\phi_{\psi}(C)}{\theta}=0
\Rightarrow \lim_{\theta \to 0}
\frac{\phi_{\theta+\psi}(C)-\phi_{\psi}(C)}{\theta}=0 \edm The
last expression says that: \bdm
\frac{d\phi_{\psi}(C)}{d\psi}=0,~~~\forall~\psi \edm i.e. the
mapping $\phi_{\psi}(C)$ is constant $\forall~\psi$. Thus it
holds, in particular, that $\phi_{\psi}(C)=\phi_{0}(C)$ or
$\widetilde{C}^{\alpha}_{\mu\nu}=C^{\alpha}_{\mu\nu}$.
\end{proof}
We have thus proven, that the finite motions induced on $\Delta$
by $X_{(i)}$ (through its integral curves), are linear
transformations of $\gamma_{\alpha\beta}$, of the form
(\ref{eq27}) with $\Lambda \in Aut(G)$. In particular, it is
deduced that the linear constraint vector fields generate inner
automorphic motions (see \cite{jantzen,ashtekar}).

We now turn out attention, to the invariant description of Bianchi
Homogeneous (B.H.) 3-Geometries. It is known that a geometry is
invariantly characterized by all its metric invariants. In 3
dimensions all metric invariants, are higher derivative curvature
invariants \cite{munoz}, and homogeneity reduces any higher
derivative curvature invariant, to a scalar combination of
$C^{\lambda}_{\alpha\beta}$,  $\gamma_{\mu\nu}$ --with the
appropriate number of $C$'s. So, it is natural to expect that
these scalar combinations, will invariantly describe a B.H.
3-geometry. Indeed, it is straightforward to check that any given
scalar combination of $C^{\lambda}_{\alpha\beta}$,
$\gamma_{\mu\nu}$, is annihilated by all $X_{(i)}$ defined in
(\ref{vectorfield}). This, in turn, implies that any such scalar
combination is constant (as a function of $\gamma_{\mu\nu}$),
along the integral curves of the $X_{(i)}$'s. This fact on account
of proposition 4, points to the following\\
\textbf{\emph{Statement}}:

\emph{Any two hexads $\gamma^{(2)}_{\alpha\beta}$,
$\gamma^{(1)}_{\alpha\beta}$, for which \underline{all scalar}
combinations of $C^{\alpha}_{\mu\nu}, \gamma_{\alpha\beta}$
coincide, are automorphically related, i.e. (\ref{eq27}) holds
with $\Lambda \in Aut(G)$.}

In order to proceed with the proof, and for latter use as well, it
is necessary to define the following scalar combinations of
$C^{\alpha}_{\mu\nu}, \gamma_{\alpha\beta}$ --which constitute a
base in the space of all scalar contractions:
\begin{subequations} \label{qdefinition}
\begin{align}
q^{1}(C^{\alpha}_{\mu\nu},
\gamma_{\alpha\beta})&=\frac{m^{\alpha\beta}\gamma_{\alpha\beta}}{\sqrt{\gamma}}\\
q^{2}(C^{\alpha}_{\mu\nu},
\gamma_{\alpha\beta})&=\frac{(m^{\alpha\beta}\gamma_{\alpha\beta})^{2}}{2\gamma}-
\frac{1}{4}C^{\alpha}_{\mu\kappa}C^{\beta}_{\nu\lambda}\gamma_{\alpha\beta}
\gamma^{\mu\nu}\gamma^{\kappa\lambda}\\
q^{3}(C^{\alpha}_{\mu\nu},
\gamma_{\alpha\beta})&=\frac{m}{\sqrt{\gamma}}
\end{align}
\end{subequations}
where $m^{\alpha\beta}$ is the symmetric second rank contravariant
tensor density (under the action of $GL(3, \Re)$ in which the
structure constants are uniquely decomposed), and $m$ its
determinant i.e.: \be \label{cdecompositon}
C^{\alpha}_{\beta\gamma}=m^{\alpha\delta}\varepsilon_{\delta\beta\gamma}+
\nu_{\beta}\delta^{\alpha}_{\gamma}-\nu_{\gamma}\delta^{\alpha}_{\beta}
\ee with $\nu_{\alpha}=\frac{1}{2}C^{\rho}_{\alpha\rho}$. At this
point the following --easily provable-- elements, must be
underlined:
\begin{itemize}
\item[$E_{1}$] Concerning the number of scalar combinations:
the number of independent $\gamma_{\alpha\beta}$'s in a $d$
dimensional space, is $N_{1}=d^{2}-\left(^{d}_{2}\right)=d(d+1)/2$
--due to its symmetry. Initially, the number of independent
structure constants, is
$N_{2}=d\left(^{d}_{2}\right)=d^{2}(d-1)/2$ --due to the
antisymmetry in its lower indices. Taking into account the number
of independent Jacobi identities, which is $
\left(^{d}_{2}\right)(d-2)=(d-2)(d-1)d/2$, one is left with
$N_{3}=N_{2}-(d-2)(d-1)d/2=(d-1)d$ independent structure
constants. But, there is also the freedom of arbitrarily choosing
$N_{4}=d^{2}$ parameters by linear mixing, i.e. the action of the
$GL(3, \Re)$. Thus, the number of independent scalars, which one
may construct from the $\gamma_{\alpha\beta}$'s and the
$C^{\alpha}_{\mu\nu}$'s, is: $N_{s}\equiv
N_{1}+N_{3}-N_{4}=(d-1)d/2$. For $d=3$, $N_{s}=3$; note that 3 is
the maximum number which is achieved only for Bianchi Type $VIII$,
$IX$. In all others, $m=0$ and, as it can be seen either by direct
calculation or from the appendix of \cite{chrissht}, the
independent scalars are less than 3; namely it is two for Type
$VI$, $VII$, $IV$ one for Type $II$, $V$ and 0 for Type $I$. In
each case, the number of the independent $q^{i}$'s equals the
number of curvature invariants.
\item[$E_{2}$] The $q^{i}$'s constitute a complete set of solutions to the
system of equations $X_{(i)}\Psi=0$ i.e. $\Psi=\Psi(q^{i})$ is the
most general solution to these equations. Since the linear
constraint vector fields, $H_{\alpha}$ are in general a subset of
the $X_{(i)}$'s, it can be inferred that the $q^{i}$'s, are
solutions to the quantum linear constraints. Except for Type
$VIII$, $IX$, where there are not extra generators, the
independent solutions to the quantum linear constraints, include
$\gamma=|\gamma_{\alpha\beta}|$ as well as other non scalar
combinations \cite{christype2,christype5}. This signals an
apparent discrepancy between the kinematics of B.H. 3-spaces
previously described, and the quantum dynamics of the (lower)
Class A Bianchi Cosmologies.
\end{itemize}

Now, to resume the line of thought for the proof of the statement,
let us define the action of $GL(3, \Re)$ on $\Delta$ and
$\mathcal{A}$. If $\Lambda^{\alpha}_{\mu},
~S^{\alpha}_{\mu}=(\Lambda^{-1})^{\alpha}_{\mu}~\in GL(3, \Re)$
then:
\begin{subequations}
\begin{align}
\widetilde{\gamma}=\phi_{\Lambda}(\gamma)&\stackrel{\textrm{def}}{\longleftrightarrow}
\widetilde{\gamma}_{\alpha\beta}=\Lambda^{\mu}_{\alpha}\Lambda^{\nu}_{\beta}\gamma_{\mu\nu}\\
\widetilde{C}=\phi_{\Lambda}(C)&\stackrel{\textrm{def}}{\longleftrightarrow}
\widetilde{C}^{\alpha}_{\mu\nu}=S^{\alpha}_{\beta}\Lambda^{\kappa}_{\mu}\Lambda^{\lambda}_{\nu}
C^{\beta}_{\kappa\lambda}
\end{align}
\end{subequations}

As it can be easily inferred from (\ref{cdecompositon}),
$\widetilde{C}=\phi_{\Lambda}(C)\Longrightarrow\widetilde{m}^{\alpha\beta}=
|S|^{-1}S^{\alpha}_{\kappa}S^{\beta}_{\lambda}m^{\kappa\lambda}$
and
$\widetilde{\nu}_{\alpha}=\Lambda^{\beta}_{\alpha}\nu_{\beta}$. It
also holds that
$\phi_{\Lambda_{1}}\circ\phi_{\Lambda_{2}}=\phi_{\Lambda_{2}\Lambda_{1}}$
and obviously, the $q^{i}$'s in (\ref{qdefinition}) satisfy the
relation: \be \label{34} q^{i}(\gamma,
C)=q^{i}(\phi_{\Lambda}(\gamma), \phi_{\Lambda}(C)) \ee This has
the important implication that, when $\Lambda^{\alpha}_{\mu}$ $\in
Aut(G)$, the form invariance of the $q^{i}$'s is guarantied by
their \underline{explicit} definition as scalar combinations of
$\gamma_{\alpha\beta}$'s and $C^{\alpha}_{\mu\nu}$'s. The
following proposition holds:
\begin{thm}
Let ~$\gamma^{(1)}_{\alpha\beta}, \gamma^{(2)}_{\alpha\beta}, ~\in
\Delta$, and $C ~\in \mathcal{A}$ be the structure constants of a
given Bianchi Type. If $q^{i}(\gamma^{(1)}, C)=q^{i}(\gamma^{(2)},
C) ~(i=1,2,3)$, then there is $\Lambda^{\alpha}_{\mu}$ such that
$\gamma^{(2)}=\phi_{\Lambda}(\gamma^{(1)})$ and
$\Lambda^{\alpha}_{\mu} ~\in Aut(G)$ i.e. $C=\phi_{\Lambda}(C)$.
\end{thm}
To prove this we need the following
\begin{lm}
If $q^{i}(I_{3}, C_{(1)})=q^{i}(I_{3}, C_{(2)}) ~(i=1,2,3)$ where
$C_{(1)}, C_{(2)} ~\in \mathcal{A}$ are two sets of structure
constants corresponding to the same Bianchi Type and $I_{3}$ is
the Identity $3\times 3$ matrix, then there exists a matrix $R
~\in SO(3)$ such that $\phi_{R}(C_{(1)})=C_{(2)}$.
\end{lm}
\begin{proof}[\textrm{Proof of the Lemma}]
We first note that the number of independent relations in Lemma's
hypothesis, equals the number of independent $q^{i}$'s and is
therefore, at most 3. We second observe that in Class A Bianchi
Types, the structure constants are characterized by the matrix
$m^{\alpha\beta}$ only, and thus the relevant numbers involved are
the (at most 3) real, non zero, eigenvalues of $m^{\alpha\beta}$.
In Bianchi Type $VIII$ and $IX$, the non vanishing eigenvalues,
are exactly 3. In conclusion, in each and every Class A Bianchi
Type, the number of independent relations in Lemma's Hypothesis,
exactly equals the number of the non vanishing eigenvalues of
matrix $m^{\alpha\beta}$.

In Class B, the null eigenvector $\nu_{\alpha}$ of
$m^{\alpha\beta}$, is also present. In this case, $q^{3}$ vanishes
identically, since rank($m$) is less than 3 and the number of
independent relations in Lemma's Hypothesis is reduced to at most
2. An apparent complication, is thus emerging for Class B Type
$VI$ and $VII$, where the independent relations are two while the
relevant numbers are 3 (the two real, non zero, eigenvalues of
$m^{\alpha\beta}$ plus the non vanishing component of
$\nu_{\alpha}$).\\
The resolution to this apparent complication, is provided by the
algebraic invariant: \bdm
\lambda\equiv\frac{C^{\tau}_{\tau\mu}C^{\chi}_{\chi\nu}I^{\mu\nu}_{3}}
{C^{\tau}_{\chi\mu}C^{\chi}_{\tau\nu}I^{\mu\nu}_{3}} \edm This
quantity, which is not meant to replace the dynamical variable
$q^{3}$, vanishes identically in Class A models, while in Class B
provides the third relation needed (see \cite{chrissht}).

Thus in every Bianchi Type, 6 numbers appear: in Class A, the 3
eigenvalues of $m^{\alpha\beta}_{(1)}$ which correspond to
$C_{(1)}$, and the 3 eigenvalues of $m^{\alpha\beta}_{(2)}$ which
correspond to $C_{(2)}$. Similarly, in Class B, the at most 2
eigenvalues of $m^{\alpha\beta}_{(1)}$ plus the third component of
its null eigenvector which correspond to $C_{(1)}$ and the at most
2 eigenvalues of $m^{\alpha\beta}_{(2)}$ plus the third component
of its null eigenvector which correspond to $C_{(2)}$. The justification
for considering only these two triplets and not --for example-- the non
diagonal components of $m^{\alpha\beta}$, lies in the fact that $m^{\alpha\beta}$
can be put in diagonal form through the action of $SO(3)$, while $\nu_{\alpha}$
will have the proper form, for being null eigenvector of $m^{\alpha\beta}$.\\
So, taking this irreducible form for both the matrix and its null
eigenvector, we have the following relations:\\
In Class A: \bdm
\begin{array}{ccc}
q^{1}(I_{3}, C_{1})=q^{1}(I_{3}, C_{2})\\
q^{2}(I_{3}, C_{1})=q^{2}(I_{3}, C_{2})\\
q^{3}(I_{3}, C_{1})=q^{3}(I_{3}, C_{2})\\
\end{array}
\edm while in Class B: \bdm
\begin{array}{ccc}
q^{1}(I_{3}, C_{1})=q^{1}(I_{3}, C_{2})\\
q^{2}(I_{3}, C_{1})=q^{2}(I_{3}, C_{2})\\
\lambda(I_{3}, C_{1})=\lambda(I_{3}, C_{2})
\end{array}
\edm

In each and every case, the corresponding system, can be easily
solved, resulting in the equality between the eigenvalues of
$m^{\alpha\beta}_{(1)}$ and $m^{\alpha\beta}_{(2)}$, as well as
$\nu_{\alpha(1)}$ and $\nu_{\alpha(2)}$. There is thus, a matrix
$R ~\in SO(3)$, such that (in matrix notation) $m_{(2)}=|R|^{-1}R~
m_{(1)}~R^{T}~~\textrm{and}\\ \nu_{(2)}=(R^{-1})^{T}\nu_{(1)}
\Longleftrightarrow C_{(2)}=\phi_{R}(C_{(1)})$ Of course, $|R|=1$
and is there, only as a reminder of the tensor density character
of $m^{\alpha\beta}$.
\end{proof}
\begin{proof}[\textrm{Proof of the Preposition 5}]
Since the matrices $\gamma^{(1)}, \gamma^{(2)}$ are positive
definite, there are $\Lambda_{(1)}, \Lambda_{(2)} ~\in GL(3, \Re)$
such that $\gamma^{(1)}=\phi_{\Lambda_{(1)}}(I_{3}),
~\gamma^{(2)}=\phi_{\Lambda_{(2)}}(I_{3})$. Let $C_{(1)}, C_{(2)}$
be defined as
$C_{(1)}=\phi_{\Lambda^{-1}_{(1)}}(C)\Longleftrightarrow
C=\phi_{\Lambda_{(1)}}(C_{(1)})$ and
$C_{(2)}=\phi_{\Lambda^{-1}_{(2)}}(C)\Longleftrightarrow
C=\phi_{\Lambda_{(2)}}(C_{(2)})$. With $C$ representing again a
given, albeit arbitrary Bianchi Type. Using the above and
(\ref{34}) we have: \bdm
\begin{array}{l}
q^{i}(\gamma^{(1)}, C)=q^{i}(\phi_{\Lambda_{(1)}}(I_{3}), \phi_{\Lambda_{(1)}}(C_{(1)}))=q^{i}(I_{3}, C_{(1)})\\
q^{i}(\gamma^{(2)}, C)=q^{i}(\phi_{\Lambda_{(2)}}(I_{3}),
\phi_{\Lambda_{(2)}}(C_{(2)})) =q^{i}(I_{3}, C_{(2)})
\end{array}
\edm The hypothesis $q^{i}(\gamma^{1}, C)=q^{i}(\gamma^{2}, C)$
translates into $q^{i}(I_{3}, C_{(1)})=q^{i}(I_{3}, C_{(2)})$
which through the lemma implies that there is $R ~\in SO(3)$ such
that $C_{(2)}=\phi_{R}(C_{(1)})$. Since $R ~\in SO(3)$ (in matrix
notation): \bdm I_{3}=\phi_{R}(I_{3})\Longrightarrow
\phi^{-1}_{\Lambda_{(2)}}(\gamma^{(2)})=\phi_{R}(\phi^{-1}_{\Lambda_{(1)}}(\gamma^{(1)}))\Longrightarrow
\gamma^{(2)}=\phi_{\Lambda_{(2)}}\circ\phi_{R}\circ\phi_{\Lambda^{-1}_{(1)}}(\gamma^{(1)})
\edm Similarly, we have: \bdm
C_{(2)}=\phi_{R}(C_{(1)})\Longrightarrow
\phi^{-1}_{\Lambda_{(2)}}(C)=\phi_{R}(\phi^{-1}_{\Lambda_{(1)}}(C)\Longrightarrow
C=\phi_{\Lambda_{(2)}}\circ\phi_{R}\circ\phi_{\Lambda^{-1}_{(1)}}(C)
\edm The above imply that the matrix
$\Lambda=\Lambda^{-1}_{(1)}R\Lambda_{(2)}$ satisfies:
$\gamma^{(2)}=\phi_{\Lambda}(\gamma^{(1)})$ and
$C=\phi_{\Lambda}(C)$ i.e. $\Lambda ~\in Aut(G)$.
\end{proof}
We have thus completed the proof of the statement that whenever
two hexads form the same multiplet $(q^{i})$, they are in
automorphic correspondence i.e. (in matrix notation): \bdm \exists
~\Lambda ~\in Aut(G):
~\gamma^{(2)}=\Lambda^{T}\gamma^{(1)}\Lambda\edm
\section{Automorphisms and the Linear Constraints}
We deem it appropriate to begin this section with a short
recalling of the main points of the quantum theory developed in
\cite{chrissht}: our starting point is the line element describing
the most general spatially homogeneous Bianchi type geometry: \be
\label{lineelement}
ds^{2}=(-N^{2}(t)+N_{\alpha}(t)N^{\alpha}(t))dt^{2}+2N_{\alpha}(t)
\sigma^{\alpha}_{i}(x)dx^{i}dt+
\gamma_{\alpha\beta}(t)\sigma^{\alpha}_{i}(x)\sigma^{\beta}_{j}(x)dx^{i}
dx^{j} \ee where $\sigma^{\alpha}_{i}$, are the invariant basis
one-forms of the homogeneous surfaces of simultaneity
$\Sigma_{t}$. Lower case Latin indices, are world (tensor) indices
and range from 1 to 3, while lower case Greek indices, number the
different basis one-forms and take values in the same range. The
exterior derivative of any basis one-form (being a two-form), is
expressible as a linear combination of any two of them, i.e.: \be
\label{2form}
d\sigma^{\alpha}=C^{\alpha}_{\beta\gamma}~\sigma^{\beta}\wedge
\sigma^{\gamma}\Leftrightarrow \sigma^{\alpha}_{i
,~j}-\sigma^{\alpha}_{j,~i}=2C^{\alpha}_{\beta\gamma}~
\sigma^{\gamma}_{i}~\sigma^{\beta}_{j} \ee The coefficients
$C^{a}_{\mu\nu}$  are, in general, functions of the point $x$.
When the space is homogeneous and admits a 3-dimensional isometry
group, there exist 3 one-forms such that the $C$'s become
independent of $x$, and are then called structure constants of the
corresponding isometry group.

Einstein's Field Equations for metric (\ref{lineelement}), are
obtained, only for the class A subgroup \cite{maccallum} (i.e.
those spaces with $C^{\alpha}_{\alpha\beta}=0$), from the
following Hamiltonian: \be \label{hamiltonian}
H=N(t)H_{0}+N^{\alpha}(t)H_{\alpha} \ee where: \be
\label{quadraticconstraint}
H_{0}=\frac{1}{2}\gamma^{-1/2}L_{\alpha\beta\mu\nu}\pi^{\alpha\beta}
\pi^{\mu\nu}+\gamma^{1/2} R \ee is the quadratic constraint, with:
\be \label{lambdarho}
\begin{array}{l}
  L_{\alpha\beta\mu\nu}=\gamma_{\alpha\mu}\gamma_{\beta\nu}+
  \gamma_{\alpha\nu}\gamma_{\beta\mu}-
\gamma_{\alpha\beta}\gamma_{\mu\nu}\\
  R=C^{\beta}_{\lambda\mu}C^{\alpha}_{\theta\tau}\gamma_{\alpha\beta}
  \gamma^{\theta\lambda}
\gamma^{\tau\mu}+2C^{\alpha}_{\beta\delta}C^{\delta}_{\nu\alpha}
\gamma^{\beta\nu}
\end{array}
\ee $\gamma$ being the determinant of $\gamma_{\alpha\beta}$, $R$
being the Ricci scalar of the slice $t =\textrm{const}$, and: \be
\label{linearconstraints}
H_{\alpha}=4C^{\mu}_{\alpha\rho}\gamma_{\beta\mu}\pi^{\beta\rho}
\ee are the linear constraints.

For Bianchi Types $VI_{0}, VII_{0}, VIII$ and $IX$, all three
$H_{\alpha}$'s are independent. Following Kucha\v{r} \& Hajicek
\cite{hajiceck}, we can quantize the system (\ref{hamiltonian})
--with the allocations (\ref{quadraticconstraint}),
(\ref{lambdarho}), (\ref{linearconstraints})-- by writing the
operator constraint equations as: \be
\label{quantumlinearconstraints}
\widehat{H}_{\alpha}\Psi=C^{\beta}_{\alpha\mu}\gamma_{\beta\nu}\frac{\partial
\Psi}{\partial \gamma_{\mu\nu}}=0 \ee \be \label{wdw}
\widehat{H}_{0}\Psi=-\frac{1}{2}\left(\Sigma^{ij}\frac{\partial^{2}\Psi}{\partial
x^{i}x^{j}}-\Sigma^{ij}\Gamma^{k}_{ij}\frac{\partial\Psi}{\partial
x^{k}}+\frac{(\mathcal{D}-2)}{4(\mathcal{D}-1)}R_{\Sigma}+\sqrt{\gamma}R\right)
\ee where $x^{i}$ are the independent solutions to
(\ref{quantumlinearconstraints}) and: \bdm
\Sigma^{ij}=\frac{\partial
x^{i}}{\partial\gamma_{\alpha\beta}}\frac{\partial
x^{j}}{\partial\gamma_{\mu\nu}}\gamma^{-1/2}L_{\alpha\beta\mu\nu}
\edm is the induced metric on the reduced configuration space.
Also, $\Gamma^{k}_{ij}, R_{\Sigma}$, are the corresponding
Christoffel symbols and Ricci scalar respectively, while
$\mathcal{D}=3$ (for details such as consistency e.t.c. see
\cite{chrissht}). The linear equations
(\ref{quantumlinearconstraints}) constitute a system of three
independent, first order, P.D.E.'s in the six variables
$\gamma_{\alpha\beta}$. These equations, by virtue of the first
class algebra satisfied by the operator constraints, admit three
independent, non-zero, solutions which can be taken to be the
combinations: \be \label{kucharvariables}
x^{1}=C^{\alpha}_{\mu\kappa}C^{\beta}_{\nu\lambda}\gamma_{\alpha\beta}
\gamma^{\mu\nu}\gamma^{\kappa\lambda}~~~~~
x^{2}=C^{\alpha}_{\beta\kappa}C^{\beta}_{\alpha\lambda}\gamma^{\kappa\lambda}~~~~~
x^{3}=\gamma \ee or any other three independent functions thereof.
These are Kucha\v{r}'s physical variables, wich solve the linear
constraints. Thus, the presence of the linear constraints at the
quantum level, implies that the state vector $\Psi$ must be an
arbitrary function of the three combinations
(\ref{kucharvariables}) or any three independent functions
thereof. This assumption is also compatible with (\ref{wdw}),
which finally becomes a P.D.E. in terms of the $x^{i}$'s (see
(2.11) of \cite{chrissht}).\\
In Type II, where only two of the three $H_{\alpha}$'s are
independent, yet another combination of $\gamma_{\alpha\beta}$'s
(except the three $x^{i}$'s in (\ref{kucharvariables})) solves
(\ref{quantumlinearconstraints}) --see first of \cite{christype2}.
In Type I, all six $\gamma_{\alpha\beta}$'s solve the identically
vanishing quantum linear constraints.

Let us now compare this theory with the purely kinematical results
of the previous section: to this end, first note that $q^{1}$,
$q^{2}$ in (\ref{qdefinition}) solve the quantum linear
constraints since, as it can be easily verified: \bdm
q^{1}=\varepsilon\sqrt{\frac{x^{1}-2x^{2}}{2}}~~~~~
q^{2}=-\frac{x^{2}}{2} \edm where
$\varepsilon=\textrm{sign}(m^{\alpha\beta}\gamma_{\alpha\beta})$
--see appendix of \cite{chrissht}.\\
In Bianchi Type $VIII$, $IX$ the existence of the non vanishing
c-number density $m$ permits us to relate $x^{3}$ to the scalar
$q^{3}=m/\sqrt{x^{3}}$; thus the grouping entailed by the quantum
Hamiltonian dynamics, is completely equivalent to that enforced by
the Kinematics of B.H. 3-spaces --described in the previous
section.\\
For Type $VI_{0}$, $VII_{0}$, $q^{3}=0$ (since $m=0$) and an
apparent discrepancy occurs: kinematically $q^{1}$, $q^{2}$, (or
equivalently $x^{1}$, $x^{2}$) invariantly and irreducibly
characterize a B.H. 3-geometry; that is any function of the
3-geometry, must necessarily and exclusively depend on $x^{1}$,
$x^{2}$. On the other hand, the quantum Hamiltonian dynamics
emanating from (\ref{hamiltonian}) allows $x^{3}=\gamma$ as a
third possible argument of the wave function which is to solve
(\ref{wdw}). The situation is getting worst when coming to the
lower Class A Types. In Type $II$, the single independent scalar
$q^{1}$, is adequate for characterizing the 3-slice while --as
explained above-- the dynamics allows $\gamma$ plus two more
combinations of $\gamma_{\alpha\beta}$'s. In Type $I$, not a
single $q^{i}$, survives while all $\gamma_{\alpha\beta}$'s are
--in principle-- candidates as arguments of the solution to the
Wheeler-DeWitt equation (\ref{wdw}).\\
The discrepancy is not of merely academic interest. Any possible
argument of the wave function other that the $q^{i}$'s (or three
independent functions thereof) is a gauge degree of freedom since
it can be affected by an appropriate A.I.D. A satisfactory
solution of the puzzle can be achieved through the usage of the
existing conditional symmetries of system (\ref{hamiltonian}). The
detailed analysis has been given for Type $I$ in the last of
\cite{ashtekar}, Type $II$ in second of \cite{christype2}, Type
$VI_{0}$, $VII_{0}$ in \cite{christype67}.\\
In the rest of this section, we give a brief outline of this idea,
and present the characteristic example of Type $V$, where a
complete matching between kinematics and Hamiltonian dynamics
occurs.

We first observe that the root of the problem lies in the
existence of the generators of the outer automorphic motions
$E_{(j)}=\varepsilon^{\sigma}_{(j)\rho}\gamma_{\sigma\tau}\frac{\partial}{\partial\gamma_{}\rho\tau}$
among the $X_{(j)}$'s: their classical counterparts
$E_{(j)}=\varepsilon^{\sigma}_{(j)\rho}\gamma_{\sigma\tau}\pi^{\rho\sigma}$
are, at first sight, absent from (\ref{hamiltonian}). As one can
easily compute, the Lie Brackets among these and the generators of
the inner automorphic motions $H_{\alpha}$'s, are: \be
\label{totalalgebra}
\{H_{\alpha},H_{\beta}\}=-\frac{1}{2}C^{\delta}_{\alpha\beta}H_{\delta}~~~
\{E_{(i)},H_{\beta}\}=-\frac{1}{2}\lambda^{\delta}_{(i)\beta}H_{\delta}~~~
\{E_{(i)},E_{(j)}\}=C^{'(k)}_{(i)(j)}E_{(k)} \ee where $\{~,~\}$
stands for the Lie Bracket and $C^{'(k)}_{(i)(j)}$ are particular
to each Bianchi Type. So, all the quantum analogues of the
$X_{(j)}$'s can be consistently imposed on the wave function: \be
X_{(j)}\Psi=0 \ee as kinematics, dictates. Then, $\Psi$ is a
function of the $q^{i}$'s only --see
Table.\\
The classical dynamics of action (\ref{hamiltonian}), provides us
some, linear in momenta, integrals of motion which are either
$E_{(j)}$'s themselves or linear combinations of some of them with
$\gamma_{\mu\nu}\pi^{\mu\nu}$ (last of \cite{ashtekar}, second of
\cite{christype2}, \cite{christype67}). Adopting the recipe that
these integrals of motion should also be turned into operators
annihilating the wave function, we achieve the desired
reconciliation between kinematics and Quantum Hamiltonian
dynamics. A very interesting feature is that the corresponding
constants of motion, are set equal to zero due to the consistency
required (Frobenius Theorem). The following general Type $V$ case,
is characteristic:

Although Type $V$ is a Class B model, a valid totally scalar
Hamiltonian has been found \cite{chrisrevd}, having the form: \be
\label{newhamiltonian} H^{c}\equiv
N_{0}H^{c}_{0}+N^{\rho}H^{0}_{\rho}=
N_{0}\left(\frac{1}{2}\Theta_{\alpha\beta\mu\nu}\pi^{\alpha\beta}\pi^{\mu\nu}+V\right)
+N^{\rho}C^{\alpha}_{\rho\beta}\gamma_{\alpha\delta}\pi^{\beta\delta}
\ee where $\Theta_{\mu\nu\rho\sigma}$ and V a 4th-rank
contravariant tensor and scalar respectively, constructed out of
the structure constants and $\gamma_{\alpha\beta}$'s. When
quantized according to Kucha\v{r} \& Hajicek, this action gives
rise to a wave function depending on 3 combinations of the
$\gamma_{\alpha\beta}$'s, namely $\Psi=\Psi(q^{2},
\frac{\gamma_{11}}{\gamma_{12}},\frac{\gamma_{12}}{\gamma_{22}})$
\cite{christype5}. Clearly, the two last arguments are gauge
degrees of freedom since --as one can see from the Table--
$q^{2}$, is the only invariant characterizing the 3-geometry under
discussion.\\
The elimination of these two degrees of freedom, is achieved by
considering the quantum analogues of the following three integrals
of motion admitted by (\ref{newhamiltonian})
$E_{(j)}=\varepsilon^{\sigma}_{(j)\rho}\gamma_{\sigma\tau}\pi^{\rho\tau}$,
with: \bdm
\begin{array}{ccc}
  \varepsilon^{\alpha}_{(1)\beta}=\left(\begin{array}{ccc}
  1 & 0 & 0 \\
  0 & -1 & 0 \\
  0 & 0 & 0
\end{array}\right) & \varepsilon^{\alpha}_{(2)\beta}=\left(\begin{array}{ccc}
  0 & 1 & 0 \\
  0 & 0 & 0 \\
  0 & 0 & 0
\end{array}\right) & \varepsilon^{\alpha}_{(3)\beta}=\left(\begin{array}{ccc}
  0 & 0 & 0 \\
  1 & 0 & 0 \\
  0 & 0 & 0
\end{array}\right)
\end{array}\edm
One can immediately recognize that these matrices are the outer
automorphisms of the Type $V$ Lie Algebra. Consequently, the
vector fields $E_{(j)}=\varepsilon^{\sigma}_{(j)\rho}
\gamma_{\sigma\tau}\frac{\partial}{\partial\gamma_{\rho\tau}}$ are
generating outer automorphic motions in the configuration space.
Turning these integrals of motion into operators imposed on
$\Psi$, i.e. demanding $E_{(j)}\Psi=M_{j}\Psi$ and utilizing the
algebra (say $C^{'k}_{ij}$) which the previous three matrices
obey, one arrives at the consistency condition
$C^{'k}_{ij}M_{k}=0$, implying that the constants of integration
$M_{k}$, should be set equal to zero. We thus retrieve all the
conditions $X_{(j)}\Psi=0$ --required by the kinematics.

So we have an example in which the dynamics completely complies
with the kinematical/geometrical results, obtained in section 2.
As we have earlier mentioned, the same situation occurs for all
Class A Types --when $E_{(j)}$'s exist. In the case of Type
$VIII$, $IX$ the Hamiltonian (\ref{hamiltonian}) is totally scalar
since, $m/\sqrt{\gamma}$ is the $q^{3}$ --$m$ being a c-number
density.
\section{Discussion}
In section 2, we first identified the particular class of G.C.T.'s
which preserve manifest homogeneity of the line element of the
generic B.H. 3-space. Their action on the configuration space
spanned by $\gamma_{\alpha\beta}$'s, is shown to be that of the
Automorphism group. The differential description of this action on
$\Delta$, leads us to the vector fields $X_{(j)}$'s. Their
characteristic solutions, the $q^{i}$'s, irreducibly and
invariantly label the 3-geometry. Thus for any given but arbitrary
Bianchi Type, points in $\Delta$, corresponding to the same
multiplet $q^{i}$, are automorphically related and thus G.C.T.
equivalent. A first conclusion concerning any possible quantum
theory of Bianchi Cosmologies, is thus reached on solely
kinematical grounds; the wave function must depend on $q^{i}$'s
only --if it is to represent the geometry and not the coordinate
system on the 3-slice.

In section 3 we first present the quantization of Hamiltonian
action (\ref{hamiltonian}) according to Kucha\v{r}'s and
Hajiceck's recipe. We see that the quantum linear constraint
vector fields $H_{\alpha}$'s corresponding to the inner
automorphisms proper invariant subgroup $InAut(G)$ of $Aut(G)$ are
among the $X_{(i)}$'s.\\
As seen from the table, for Types $VIII$,
$IX$ there are no outer-automorphisms and the three $x^{i}$'s are
in one-to-one correspondence to the three $q^{i}$'s (essentially
the three independent curvature invariants). For all other Class A
Types, there is always an outer-automorphism with non-vanishing
trace; the corresponding generator in configuration space $\Delta$
does not (weakly) commute with the quadratic constraint
(\ref{quadraticconstraint}) nor does its corresponding quantum
analogue commute with (\ref{wdw}). Thus, for the lower Class A
Types, the wave functions emanating from action
(\ref{hamiltonian}), depend on the curvature invariants and on
$\gamma$ despite that $q^{3}=0$; these wave functions will
therefore not be G.C.T. invariant, since $\gamma$ can be changed
to anything we like by an A.I.D. This result seems to justify (for
these types) the claim made by some authors, that $\gamma$ should
be considered as time variable and thus frozen out
\cite{schirmer}. One may say that for the lower Class A Types  the
grouping dictated by the quantum theory, resulting from action
(\ref{hamiltonian}), is overcomplete: although any two hexads
forming the same $x^{1}, x^{2}$ are geometrically identifiable
(since they correspond to G.C.T. related spatial line-elements),
the theory requires that $x^{3}=\gamma$ be also the same in order
to consider these two hexads as equivalent. At first sight, this
may be seen as a defect of the classical action
(\ref{hamiltonian}); although it reproduces Einstein's Equations
for (Class A) spatially homogeneous spacetimes, it does not
correctly reflects the full covariances of these equations.
However, as is explained in
\cite{christype2,christype5,christype67,ashtekar}, the conditional
symmetries of this action rectify this defect: for Bianchi Types
other than $VIII$, $IX$, there are extra, linear in momenta,
integrals of motion --say
$E_{(i)}=\varepsilon^{\alpha}_{(i)\rho}\gamma_{\alpha\sigma}\pi^{\rho\sigma}$--
corresponding to the outer automorphisms subgroup of $Aut(G)$. It
is shown how the quantum analogues of these $E_{(i)}$'s can serve
to satisfactorily remove this discrepancy. Their imposition as
additional conditions restricting the wave function, results in
forcing it to depend on $q^{i}$'s only. A noteworthy feature of
this procedure is that, at the quantum level, the consistency
requirement of these extra conditions leads to setting zero, the
classical constants of integration (which are non essential) --as
the particular example of Type $V$, exhibits.

Another important consequence of the results in section 2, is the
conclusion that a Homogeneous 3-Geometries are completely
characterized by their curvature invariants: indeed, as it is well
known, in 3 dimensions all metric invariants are higher derivative
curvature invariants \cite{munoz}; but the homogeneity of the
space reduces any higher derivative curvature invariant to a
scalar combination of $C^{\alpha}_{\mu\nu}, \gamma_{\alpha\beta}$
with the appropriate number of $C$'s. Thus any two distinct
Homogeneous 3-Geometries must differ by at least one curvature
invariant, i.e. by at least one $q^{i}$; and vice versa, any two
Homogeneous 3-metrics for which all curvature invariants (i.e. all
$q^{i}$'s) coincide, are necessarily G.C.T. related and thus
represent the same 3-Geometry.

Last but not least, we would like to underline that the
partitioning of the Automorphism Group in Inner and Outer
Subgroups, which quantum theory seems to favour, does have a
classical analogue: the inner automorphism parameters, represent
genuine `'gauge`' degrees of freedom (i.e. can be allowed to be
arbitrary functions of time) --see 4th of \cite{jantzen}--, while
the outer automorphism parameters, are rigid symmetries --3rd of
\cite{jantzen}.

\vspace{1.5cm} \large{\textbf{Acknowledgements}}\\ The authors
wish to express their appreciation for the referee's critical
comments on an earlier version of the manuscript, which helped
them to present a more clear version of the essence of this
work.\\
One of us (G. O. Papadopoulos) is currently a scholar of the Greek
State Scholarships Foundation (I.K.Y.) and acknowledges the
relevant financial support.
\newpage
\begin{center}
\large{\textbf{Table}}
\end{center}
\begin{center}
\begin{tabular}{|c|c|c|c|c|c|}
\hline \hline
Type & Generators & \# of Indep. &  \# of Indep. &  \# of Indep. &   \# of Indep.\\
         & $\lambda^{\alpha}_{(i)\beta}$ & Parameters & $H_{\alpha}$'s & $E_{\alpha}$'s & $q^{i}$'s\\
\hline \hline
I & $\left(\begin{array}{ccc}
p_{1} & p_{2} & p_{3}\\
p_{4} & p_{5} & p_{6}\\
p_{7} & p_{8} & p_{9}
\end{array}\right)$ & 9 & 0 & 0 & 0 \\
\hline
II & $\left(\begin{array}{ccc}
p_{3}+p_{6} & p_{1} & p_{2}\\
0 & p_{3} & p_{4}\\
0 & p_{5} & p_{6}
\end{array}\right)$ & 6 & 2 & 3 & 1\\
\hline
III & $\left(\begin{array}{ccc}
p_{1} & p_{2} & p_{3}\\
p_{2} & p_{1} & p_{4}\\
0 & 0 & 0
\end{array}\right)$ & 4 & 2 & 2 & 2\\
\hline
IV & $\left(\begin{array}{ccc}
p_{1} & p_{2} & p_{3}\\
0 & p_{1} & p_{4}\\
0 & 0 & 0
\end{array}\right)$ & 4 & 3 & 1 & 2\\
\hline
V & $\left(\begin{array}{ccc}
p_{1} & p_{2} & p_{3}\\
p_{4} & p_{5} & p_{6}\\
0 & 0 & 0
\end{array}\right)$ & 6 & 3 & 2 & 1\\
\hline
VI & $\left(\begin{array}{ccc}
p_{1} & p_{2} & p_{3}\\
p_{2} & p_{1} & p_{4}\\
0 & 0 & 0
\end{array}\right)$ & 4 & 3 & 1 & 2\\
\hline
VII & $\left(\begin{array}{ccc}
p_{1} & p_{2} & p_{3}\\
-p_{2} & p_{1} & p_{4}\\
0 & 0 & 0
\end{array}\right)$ & 4 & 3 & 1 & 2\\
\hline
VIII & $\left(\begin{array}{ccc}
0 & p_{1} & p_{2}\\
p_{1} & 0 & p_{3}\\
p_{2} & -p_{3} & 0
\end{array}\right)$ & 3 & 3 & 0 & 3\\
\hline
XI & $\left(\begin{array}{ccc}
0 & p_{1} & p_{2}\\
-p_{1} & 0 & p_{3}\\
-p_{2} & -p_{3} & 0
\end{array}\right)$ & 3 & 3 & 0 & 3\\
\hline
\end{tabular}
\end{center}

\vspace{1cm}
\emph{Notes}:
\begin{itemize}
\item[$N_{1}$] \emph{The number of the independent $q^{i}$'s equals the number of the independent curvature invariants.}
\item[$N_{2}$] \emph{Type III, is characterized by the condition $h=\pm 1$, while Type VI, by the condition
$h\neq (0,\pm 1)$.}
\end{itemize}

\end{document}